\documentclass[slac_one]{revtex4}
\usepackage{graphicx}
\usepackage{fancyhdr}
\def\[{\left [} 
\def\]{\right ]} 
\def\({\left (} 
\def\){\right )}

\newcommand{\lbr}{\left\{} 
\newcommand{\rbr}{\right\}} 
\newcommand{\oline}[1]{\overline{#1}}

\newcommand{\wtd}[1]{\widetilde{#1}} 

\newcommand{\hc}       {\mathrm{\; h.c. \;}}

\newcommand{\gappeq}{\mathrel{\rlap {\raise.5ex\hbox{$>$}} 
{\lower.5ex\hbox{$\sim$}}}} 
\newcommand{\lappeq}{\mathrel{\rlap{\raise.5ex\hbox{$<$}} 
{\lower.5ex\hbox{$\sim$}}}}

\def\Eisen{G_{2}\(t,\bar{t}\)}

\def\Eisen{G_{2}\(t,\bar{t}\)}

\begin{document}
\rightline{LPT--Orsay 05/33}
\rightline{DESY 05-121}

\title{{\small{2005 International Linear Collider Workshop - Stanford,
U.S.A.   }}\\ 
\vspace{12pt}

Astroparticle and Collider Physics as complementary sources for the study of string motivated supergravity models} 

%

\author{Y. Mambrini}
\affiliation{LPT, 91400 Orsay, France -- DESY, 22607 Hamburg, Germany}

\vspace{-1truecm}

\begin{abstract}
We provide a study of the phenomenology of 
heterotic orbifold compactifications scenarii within the context of
supergravity effective theories. Our investigation focuses on
those models where the soft Lagrangian is dominated by loop
contributions to the various soft supersymmetry breaking
parameters, giving a mixed anomaly-gravity mediation model. 
We consider the pattern
of masses that are governed by these soft terms and investigate
the implications of certain indirect constraints on supersymmetric
models. In this framework, we point out how the complementarity between 
direct and indirect detection of a neutralino Dark Matter, and futur
accelerator prospects can reduce considerably the parameter space of such 
models
\end{abstract}

\maketitle

\thispagestyle{fancy}


\section{INTRODUCTION} 

One of the most crucial and difficult tasks of string phenomenologists is now to 
make, and keep, contact between the high energy theory, and the low energy world. 
For that, we need to consider a superstring theory which yields in four dimensions, 
the Standard Model gauge group,  three generations of quarks, and a consistent 
mechanism of SUSY breaking. Our analysis have relies on 
orbifold compactifications of the heterotic string within the context of 
supergravity effective theories. More specifically, we concentrate on those 
models where the 
action is dominated by one loop order contributions to soft breaking terms. 
Recently, all one loop order contributions have been calculated 
\cite{BiGaNe01}. The key point of such models is the non universality of
supersymmetry breaking term which is a  
consequence of the beta--function appearing in the superconformal anomalies. 
This non universality gives a specific phenomenology in the gaugino 
and the scalar sectors, modifying the predictions coming from Msugra. 
In fact, these string--motivated models show new behavior that interpolates 
between the phenomenology of unified supergravity models (Msugra) and models 
dominated by the superconformal anomalies (AMSB). The constraints arising from 
accelerator physics, and dark matter aspects have been already studied \cite{Bin1}.
The prospect of direct detection, \cite{BGWdirect} and
indirect from the galactic center \cite{BGWindirect} have been recently published. 
 It becomes interesting now, to see in which sense 
experimental limits on supersymmetric particles  will be able to
 bring us informations, or even to rule out some of these models, taking 
into account the complementarity between accelerator physics (LEPII, future
LC) and astroparticle.

\section{THEORETICAL FRAMEWORK}

Our phenomenological study is based on orbifold compactifications of the 
weakly--coupled heterotic string, where we distinguish two regimes.
In the first one, SUSY breaking is driven by the compactification
 moduli $T$, whose vacuum expectation values determine the size of the 
 compact manifold. In the second one, it is the dilaton field $S$, whose vacuum 
 expectation value determines the magnitude of the (unified) coupling constant 
 $g_{\mathrm{STR}}$ at the string scale, that transmits, via its auxiliary fields,
  SUSY breaking. We work in the context of models in which 
  string nonperturbative corrections to the Kahler potential act to stabilize the
   dilaton in the presence of gaugino condensation
\cite{BiGaWu96}. The origins of breaking terms are diverse. 
Some coming from the superconformal anomalies are non--universal (proportional to
 the
 beta-- function of the $SU(3) \times SU(2) \times U(1)$ groups) some are 
 independent of the gauge group considered (Green--Schwarz counterterm, $vev$ 
 of the condensate). This interplay between universality and non--universality 
 gives a rich new phenomenology, and indicates new trends in the
  search of supersymmetric particles in accelerator or astroparticle physics. 
\subsection{The moduli dominated scenario}
In the moduli dominated scenario, the supersymmetric susy breaking terms can 
be written\cite{BiGaNe01}
\begin{eqnarray}
M_a &=& \frac{g_{a}^{2}\(\mu\)}{2} \lbr 2
 \[ \frac{\delta_{\mathrm{GS}}}{16\pi^{2}} + b_{a}
\]\Eisen F^{T} + \frac{2}{3}b_{a}\oline{M} \rbr, \label{modsoftgaugi}\\
A_{ijk}&=& - \frac{1}{3} \gamma_{i}\oline{M} - p \gamma_{i} \Eisen
F^{T} + {\rm cyclic}(ijk), \\ M_{i}^{2} &=& (1-p)\gamma_i
\frac{|M|^2}{9}. \label{modsoftscal}
\end{eqnarray}

where $b_a$ is the one loop beta--function coefficient of the
$SU(3)\times SU(2)\times U(1)$ gauge coupling $g_{a=1,2,3}$. 
The field $M$ is the auxiliary field of the supergravity multiplet 
related to the gravitino mass by

\begin{equation}
M_{3/2}=-\frac{1}{3}\langle \overline{M} \rangle.
\end{equation}

We clearly see in these formulae the mixing between universal term 
and non--universal ones. Moreover, scalar mass terms are coming with a 
loop suppression factor $\gamma_i$, and the gaugino mass breaking terms 
have a
 universal compensation coming from the Green--Schwarz counterterm 
 (appearing in order to cancel anomalies) that 
 can give high value to the chargino or neutralino masses. To sum up, this 
 regime gives light scalars and relatively heavy gauginos, whose nature 
 depends completely on the value of $\delta_{\mathrm{GS}}$.

\subsection{The dilaton dominated scenario}

In this region of parameter space, we can express the soft SUSY breaking
terms as

\begin{eqnarray}
M_{a}&=&\frac{g_{a}^{2}\(\mu\)}{2} \lbr  \frac{2}{3}b_{a}\oline{M}
+\[ 1 - 2 b_{a}' k_s \] F^{S} \rbr \label{dilatsoftgaugi}\\ A_{ijk} &=&
-\frac{k_s}{3}F^S - \frac{1}{3} \gamma_{i}\oline{M} +
\tilde{\gamma}_{i} F^{S} \lbr \ln(\mu_{\mathrm{PV}}^{2}/\mu_R^2)
-p\ln\[(t+\bar{t}) |\eta(t)|^4\] \rbr + (ijk) 
 \\ M_{i}^{2} &=& \frac{|M|^2}{9}
 \[ 1 + \gamma_i
-\(\sum_{a}\gamma_{i}^{a} -2\sum_{jk}\gamma_{i}^{jk}\) \(
\ln(\mu_{\mathrm{PV}}^{2}/\mu_R^2) -p\ln\[(t+\bar{t}) |\eta(t)|^4\] \) \]
\nonumber
 \\
 & &+ \lbr
 \wtd{\gamma}_{i}\frac{MF^S}{6}+\hc \rbr , \label{dilatsoftscal}
\end{eqnarray}

\noindent
with

\begin{equation}
F^{S} = 3
\frac{\frac{2}{3}b_{+}}{1-\frac{2}{3}b_{+}K_{s}} M_{3/2}.
\label{FS}
\end{equation}

\noindent
with $b_+$ being the largest beta--function coefficient among the condensing
gauge groups of the hidden sector, $k_s$ the derivative in $S$ 
of the Khaler potential and $p_i$ the Pauli--Villars weigths of the regulator fields.

The phenomenology of the dilaton dominated scenario is completly different
from the moduli dominated one. If we look at (\ref{dilatsoftscal}) and 
(\ref{dilatsoftgaugi}),
 it
is clear that we are in a domain of heavy squarks and sleptons (of the order
of the gravitino scale) and light gaugino masses, directed by the dilaton
auxiliary field $vev's$. Indeed, the beta--functions $b_a$ are of the order 
of $10^{-2}$,
which will not be competitive compared to the $F$ term of the dilaton in
(\ref{dilatsoftgaugi}). In fact, if we look more clearly at
(\ref{FS}), for not so big values of $b_+$, we can consider that $F^S$ has a
 linear evolution as a function of $b_+$. Increasing $b_+$ means approaching
 the universal case for the gaugino sector (and the scalar one, driven by 
 $M_{3/2}$).

\section{ASTRO-PHENOMENOLOGICAL ASPECTS OF THE MODELS}

In the specific context of the class of string models that we have considered,
we have seen in \cite{BGWdirect} that the prediction regarding dark matter are
strikingly different according to the type of supersymmetry breaking considered. 
In the case of moduli domination, one does not expect any signal in the forthcoming
direct or indirect (neutrino) detection experiments. On the other hand, these
experiments should not miss the neutralino signal in the case of dilaton domination.
Thus, the detection of dark matter or the absence of detection may give key
 information on the nature of supersymmetry breaking, at least in the context 
of this given class of models. 

We have also studied in \cite{BGWindirect} gamma--ray and synchrotron radiation
emission from the Galactic center in this context. Typically, as it is the 
case for direct detection, models in the dilaton dominated SUSY breaking scenario
 lead to a higher annihilation rate than the moduli scenario. Concerning the
 continuum gamma--ray flux, both scenarios are within the reach of the
experimental sensitivities of GLAST and HESS for a NFW halo profile. For
the same profile, the gamma--ray line signal is supressed and beyond the 
experimental sensitivitiy. The synchrotron emission is too low to be constrained 
by experiments even with a more cuspy profile.

Obviously, there are connections between these results and the detection of the
LSP at colliders. Small direct detection cross sections or small indirect 
detection fluxes are obviously correlated with small production cross sections at
 colliders. In any case, it is interesting for collider searches to note the
characteristics of the regions that satisfy the criterion of satisfactory relic
density. For moduli domination, we have identified two regions of interest : 
one where $m_{\chi}\sim m_{\chi^+_1}\sim m_{\chi^0_2}$ through the bino
and wino content of the LSP (for sufficiently large value of tan$\beta$), and
the other one close to the stau LSP region where 
$m_{\tilde \tau_1} \sim m_{\chi}$. In the case of dilaton domination, the
 cosmologically interesting region corresponds to a LSP with a proper higgsino
 content. Furthermore, the
parameter space being closed ($b_+$ has a maximum value of $\sim$ 0.57),
this case gives an upper bound on neutralino mass of 
$m_{\chi}\lesssim 1500$ GeV.

\begin{figure*}[t]
\centering
\includegraphics[width=195mm,angle=90]{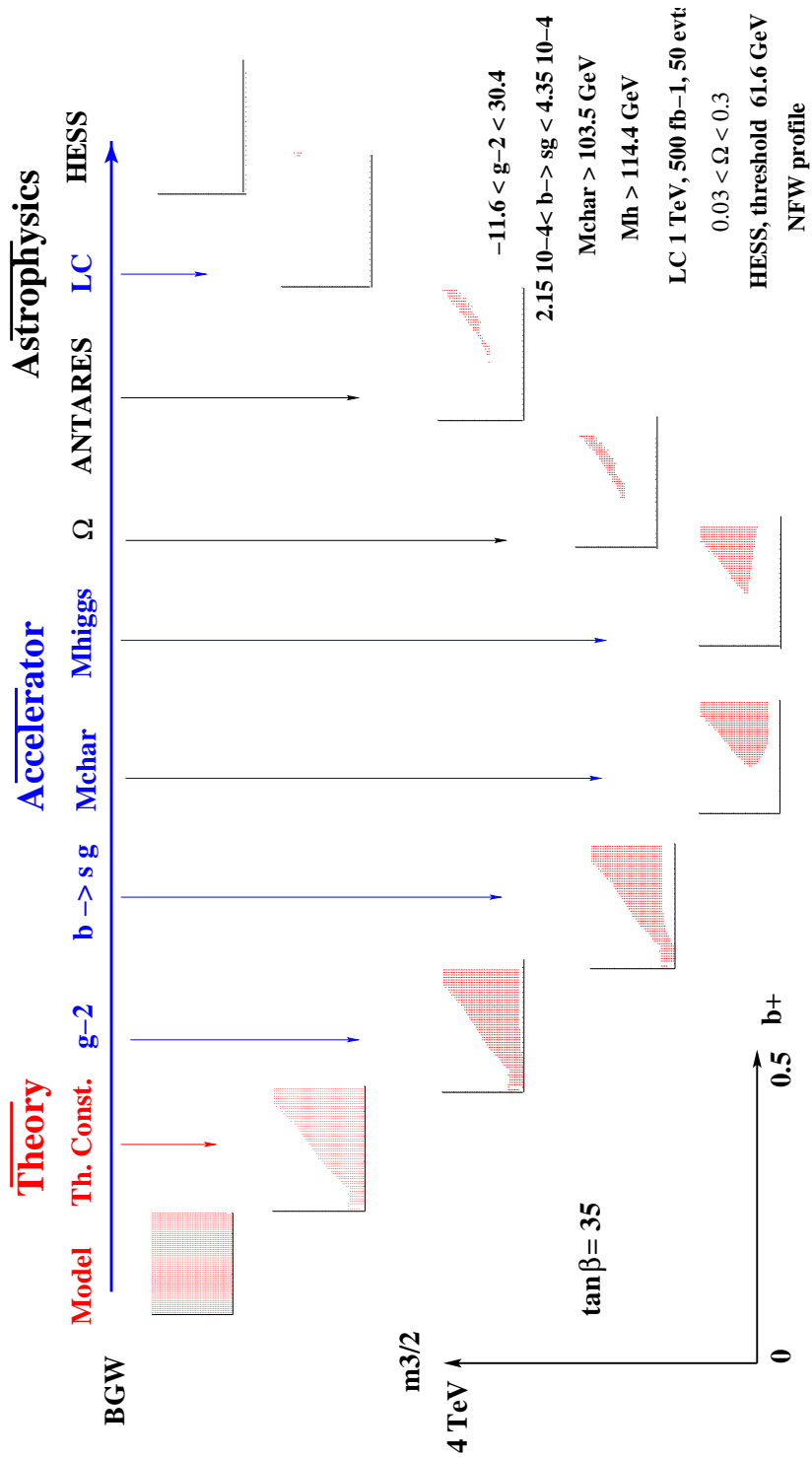}
\caption{Example of restriction of the ($b_+,~m_{3/2}$) parameter space 
in BGW model for tan$\beta=35$. We applied here the accelerator constraints coming
from LEPII on chargino and higgs mass, on the $g-2$ and $b \rightarrow s \gamma$
rare processes, indirect detection from SUN by ANTARES, from the Galactic
center from HESS and a future 1 TeV linear collider with a luminosity
of 500 $fb^{-1}$ and 50 events as a discovery in the chargino sector.
All the points appearing in the ($b_+,~m_{3/2}$) plane respect all the constraints
for present experiment, and still survive if there is no observation
from future experiment (LC, ANTARES, HESS).} 
\label{Figure}
\end{figure*}

We have sum up all these constraints and prospects using the complementarity
between accelerator physics and astroparticle in the ($b_+,m_{3/2}$) plane
in Fig. \ref{Figure} for tan$\beta=35$. We clearly see that all the parameter
space can be excluded if no discovery is made in the next generation
of collider (Linear Collider of 1 TeV) or astroparticle experiments
(indirect detection of neutrino from sun, ANTARES). This methodology
can of course be applied in any class of string inspired model, especially
the recently developped KKLT set-up \cite{KKLTpheno}.

\begin{acknowledgments}
The author wish to thank all his collaborators for this work. He also
wants to warmly thank P. Zerwas for sharing his incredible knowledge and
enthousiasm, 
and the DESY theory group for their scientific and financial supports.
\end{acknowledgments}

\end{document}